\documentclass[twocolumn,tighten]{aastex631}
\usepackage{mathrsfs}

\usepackage[dvipsnames]{xcolor}

\begin{document}

\title{Predicting Stellar Parameters of Massive Stars from Light Curves with Machine Learning}

\author[0000-0002-2905-9239]{Rachel C. Zhang}
\affiliation{Flatiron Institute Center for Computational Astrophysics, New York, NY 10010, USA}
\affiliation{Center for Interdisciplinary Exploration \& Research in Astrophysics (CIERA), Northwestern University, Evanston, IL 60208, USA}
\affiliation{Department of Physics \& Astronomy, Northwestern University, Evanston, IL 60208, USA}

\author[0000-0001-8432-7788]{Kaze W. K. Wong}
\affiliation{Flatiron Institute Center for Computational Astrophysics, New York, NY 10010, USA}
\affiliation{Department of Applied Mathematics and Statistics, Johns Hopkins University, Baltimore, Maryland, 21218}

\author[0000-0002-9296-8259]{Gonzalo Holgado}
\affiliation{Instituto de Astrofísica de Canarias, Avenida Vía Láctea, E-38205, La
Laguna, Tenerife, Spain}
\affiliation{Universidad de La Laguna, Dpto. Astrofísica, E-38206, La Laguna, Tenerife, Spain.}

\author[0000-0002-8171-8596]{Matteo Cantiello}
\affiliation{Flatiron Institute Center for Computational Astrophysics, New York, NY 10010, USA}
\affiliation{Department of Astrophysical Sciences, Princeton University, Princeton, NJ, 08544, USA}

\begin{abstract}
High-resolution spectroscopic measurements of OB stars are important for understanding processes like stellar evolution, but require labor-intensive observations. In contrast, photometric missions like the Transiting Exoplanet Survey Satellite (TESS) can monitor hundreds of thousands of stars with a range of temporal resolutions, but do not provide such detailed measurements. With surveys like the Legacy Survey of Space and Time promising unprecedented photometric coverage over the next ten years, it is increasingly important to develop methods that connect large-scale time-series photometry with the detailed stellar parameter measurements typically derived from spectroscopy. In this paper, we test whether machine learning can recover such parameters by combining TESS light curves with spectroscopic measurements from the IACOB project, using a sample of 285 light curves from 106 unique O stars. Using both multilayer perceptrons and convolutional neural networks, we demonstrate that (1) O star light curves contain sufficient information to meaningfully infer stellar parameters and (2) periodograms derived from light curves capture substantially more information than previously identified correlation parameters. Our best model achieves moderate success in predicting both spectroscopic luminosity ($R^2 = 0.641_{-0.167}^{+0.107}$) and effective temperature ($R^2 = 0.443_{-0.234}^{+0.056}$), key stellar parameters for determining positions of stars on the spectroscopic Hertzsprung–Russell diagram, despite the small dataset size. Further progress will require expanded datasets of matched photometric and spectroscopic observations.

\end{abstract}


\section{Introduction} \label{sec:intro}

The Hertzsprung-Russell (H-R) diagram \citep{hertzsprung1905, russell1919} has been used in astrophysics for over a century, illustrating how a star's absolute luminosity and effective temperature can distinguish between types of stars. In recent years, large-scale surveys have transformed our ability to construct H-R diagrams, with Gaia providing the most complete diagram to date, encompassing millions of stars \citep{gaia2023}. The spectroscopic H-R diagram \citep{langer2014} plots effective temperature against spectroscopic luminosity, as opposed to absolute luminosity. Spectroscopic luminosity is defined as:
\begin{equation}
    \label{eq1}
    \mathscr{L} \equiv {\rm T}^4_{\rm eff}/g,
\end{equation} 
where $T_{\rm eff}$ is effective temperature and $g$ is surface gravity, fundamental stellar parameters.

Currently, fundamental stellar parameters such as $T_{\rm eff}$ and $g$ are acquired through labor-intensive spectroscopic observations. Notably, the IACOB project, which aimed to observe the largest possible number of massive stars, achieved just over 1000 stars in its first 12 years using the 2.56-m Nordic Optical Telescope, the 1.2-m Mercator telescope, and the 2.2-m ESO telescope \citep{simondiaz2011, simondiaz2015, simondiaz2020}. In contrast, time-series photometry enables the monitoring of vastly larger samples of stars. For example, the Transiting Exoplanet Survey Satellite (TESS), launched in 2018, allows for the collection of light curve data from over 200,000 stars at various temporal resolutions ranging from 20 seconds to 30 minutes \citep{ricker2015}. Likewise, the Vera C. Rubin Observatory’s Legacy Survey of Space and Time (LSST), which began operations in 2025, will repeatedly image the southern sky over a 10-year baseline, constructing light curves for millions of stars \citep{ivezic2019}. Although these time-series photometric data provide rich temporal information, they do not directly yield key stellar parameters such as effective temperature, surface gravity, or spectroscopic luminosity. This growing imbalance underscores the importance of developing methods to extract stellar information directly from large-scale time-series photometry.

OB stars, being among the most luminous stellar objects, are represented in high-resolution spectroscopic surveys, such as in the IACOB project. Despite being one of the least understood stellar classes, OB stars play a fundamental role in driving galactic evolution through their intense radiation fields, powerful stellar winds, and eventual supernova explosions \citep{leitherer1994, hopkins2012}. They are primary sources of momentum and energy input into the interstellar medium, and through their deaths, they contribute significantly to galactic chemical enrichment while leaving behind compact objects such as neutron stars and black holes that can produce gravitational waves through mergers (see review by \citet{langer2012}). Additionally, OB stars exhibit rich asteroseismological behavior. Their distinct variability patterns, driven by both pressure and gravity modes, can be directly captured through time-series photometry, providing probes of their internal structure (see \citet{aerts2021} for a comprehensive review). The physical connection between this variability and stellar evolutionary state makes OB stars particularly well-suited for approaches that aim to infer stellar parameters from light curve data.

Prior studies have shown that massive star light curves may carry information about fundamental stellar parameters. \citet{blomme2011} first discussed that the red noise components of three O star samples seemed to potentially contain stellar information and speculated on the physical mechanism. In this context, ``red noise'' refers to the stochastic low-frequency variability observed in massive stars. Subsequent work with larger samples from K2 and TESS \citep{bowman2019, bowman2020, anders2023} further demonstrated that the fitted red noise components correlated with stellar parameters, such as effective temperature and spectroscopic luminosity. Hence, this red noise is thought to arise from physical processes such as internal gravity waves launched by core convection \citep{bowman2023}, waves and/or granulation generated by subsurface convection \citep{cantiello2021, schultz2022}, or wind instability \citep{krticka2021}, and therefore may encode meaningful information about stellar structure and evolution. However, these studies primarily established associations rather than predictive models, leaving open the question of how fully such red noise can be leveraged to understand massive star processes.

\begin{figure*}
    \centering
    \includegraphics[width=\linewidth]{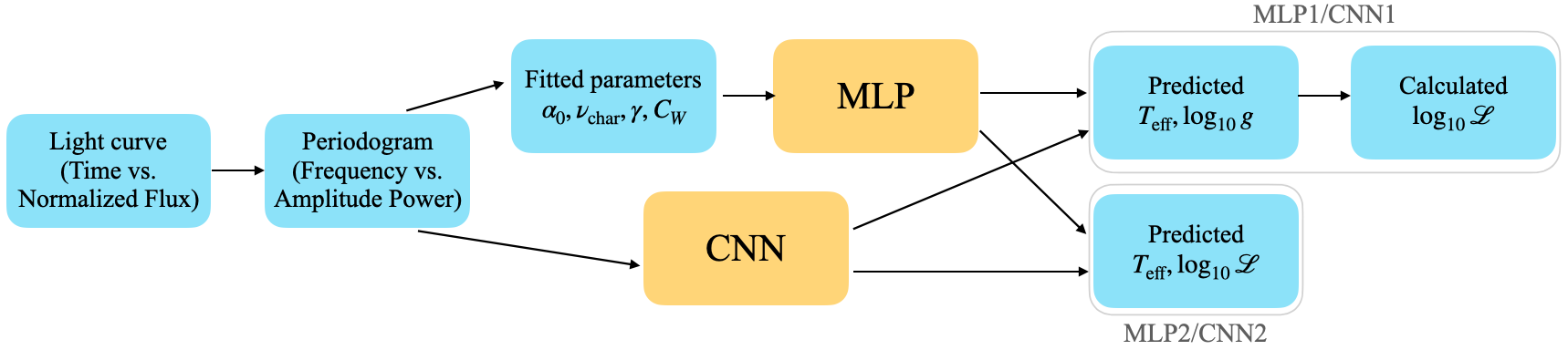}
    \caption{Workflow setup to perform each of the prediction tasks. This flowchart visually differentiates the two main models tested (MLP, CNN), highlighting how each processes inputs to predict stellar parameters. Models trained to directly predict $T_{\rm eff}$ and $\log_{10} g$ are denoted with a ``1'' (e.g. MLP1, CNN1), while models trained to directly predict $T_{\rm eff}$ and $\log_{10} \mathscr{L}$ are labeled with a ``2'' (e.g. MLP2, CNN2).}
    \label{fig1}
\end{figure*}

Independently, numerous studies have demonstrated the predictive power of stellar light curves across a variety of contexts and datasets. For classification tasks, deep learning architectures have shown particular promise, with convolutional neural networks (CNNs) effectively capturing both spatial and temporal patterns in astronomical time-series data \citep{pasquet2019, neira2020, cui2024}. Beyond classification, light curves have also proven valuable for regression tasks aimed at predicting stellar parameters. Both traditional algorithms and modern deep learning approaches have successfully extracted physical parameters such as rotation period and surface gravity directly from light curves in stars cooler than O and B stellar types \citep{miller2015, blancato2022, kamai2025}. Motivated by the success of this prior work and by evidence that the red noise component of massive star light curves encodes stellar information, we investigate whether similar strategies can be applied to massive stars.

Our work builds directly on this prior research by combining light curve data from TESS with spectroscopically derived stellar parameters from the IACOB project. We pursue two main goals: (1) testing whether massive star light curves can be used to predict stellar parameters, and (2) evaluating whether the fitted red noise parameters identified by \citet{blomme2011, bowman2019} capture all of the relevant information, or if more detailed frequency-domain representations provide additional predictive power. We demonstrate that O star light curves do contain sufficient information to predict stellar parameters, with performance strongly dependent on how the light curve is represented. The dominant red noise components contain some predictive ability, but models that incorporate the full frequency-domain representation achieve substantially higher accuracy, demonstrating that frequency-domain representations of light curves contain more stellar information than fitted red noise parameters alone.

\section{Methods}
\label{methods}

\subsection{Data}
\label{sec:data}
Our analysis uses datasets derived from NASA's TESS mission and the IACOB spectroscopic database. We downloaded 2-minute cadence light curves from TESS (600-1040 nm; \citet{vanderspek2018}) through the TESS Bulk Downloads webpage (\url{https://archive.stsci.edu/tess/bulk_downloads.html}). From these, we used the Pre-search Data Conditioned Simple Aperture Photometry (PDCSAP) flux values, which correct for systematic instrumental variations. To identify OBA-type stars, we used the catalog from \citet{ijspeert2021}, who applied J-H and J-K color cuts to the TESS Input Catalogue \citep{stassun2019}. Their analysis identified 189,981 OBA-like star candidates, of which 91,193 had available light curves. Cross-matching these candidates with our downloaded TESS data yielded 33,169 high-mass OBA-like star light curves.

The IACOB database currently contains high-resolution spectra of O and B-type stars from the HERMES \citep{raskin2011} and FIES \citep{telting2014} spectrographs. For O stars with at least one high-resolution spectrum, we used the stellar parameters derived following the methodologies outlined in \cite{holgado2018}. This approach involves using the {\sc iacob-broad} \citep{Simon-Diaz2014} tool to measure line-broadening parameters, such as the projected rotational velocity {$v$\,sin\,$i$} and macroturbulence, through a combined Fourier transform and goodness-of-fit analysis. Subsequently, the {\sc iacob-gbat} \citep{Simon-Diaz2011} tool and a grid of FASTWIND models \citep{puls2005} is employed to determine atmospheric parameters, including effective temperature, gravity, helium abundance, among others. A key strength of this methodology is that it ensures a homogeneous and consistent analysis across many O stars while retaining a semi-automatic component. Despite the high level of automation, a final visual inspection is
required to validate the results and account for peculiarities that classic automated algorithms might overlook. After applying this full process, we obtained a sample of 257 O-type stars with detailed stellar parameter information from IACOB. The complete IACOB catalog will be available at \url{https://research.iac.es/proyecto/iacob/iacobcat/}.

We then combined TESS light curves with their corresponding stellar parameters derived from IACOB by cross-matching their TIC IDs, producing a dataset restricted to stars present in both datasets. After removing stars with any $\rm NaN$ flux values, our final regression dataset comprises 285 O star light curves from TESS with corresponding labeled stellar information from IACOB. Since TESS observes stars across multiple sectors, the 285 light curves correspond to 106 unique stars, with each sector treated as an independent observation. The 106 unique O stars used comprise 19 early ($\leq \rm O6$), 53 mid ($\rm O6 <x \leq O8.5$), and 34 late ($\geq \rm O9$) type stars, with luminosity classes distributed as 53 dwarfs (IV -- V), 16 giants (III), and 37 supergiants (I -- II). Each star is characterized by measured stellar parameters: effective temperature ($T_{\rm eff}$) and surface gravity ($\log_{10} g$), and a derived spectroscopic luminosity ($\log_{10} \mathscr{L}$) defined by Equation \ref{eq1}.

\subsection{Machine learning approach}
\label{sec:mlapproach}

\begin{deluxetable*}{lcccc}[!htb]
\tablecaption{Best-performing architecture details for each prediction task.
\label{table1}}
\tablehead{
\colhead{Model} & \colhead{Input} & \colhead{Architecture} & \colhead{Output} & \colhead{\# of trainable parameters}
}
\startdata
MLP1 & \parbox[t]{2.5cm}{4 parameters ($\alpha_0$, $\nu_{\rm char}$, $\gamma$, $C_W$)} & [256, 512, 128] & $T_{\rm eff}$, $\log_{10} g$ & 198,786\\
MLP2 & \parbox[t]{2.5cm}{4 parameters ($\alpha_0$, $\nu_{\rm char}$, $\gamma$, $C_W$)} & [512, 256, 128] & $T_{\rm eff}$, $\log_{10} \mathscr{L}$ & 167,042 \\
CNN1 & Full periodogram & \parbox[t]{4.5cm}{3 Conv. layers (64 channels) + 1 FC layer (128), Max pooling} & $T_{\rm eff}$, $\log_{10} g$ & 6,439,810 \\
CNN2 & Full periodogram & \parbox[t]{4.5cm}{3 Conv. layers (64 channels) + 1 FC layer (128), Max pooling} & $T_{\rm eff}$, $\log_{10} \mathscr{L}$ & 6,439,810 \\
\enddata
\end{deluxetable*}


Our goal is to predict stellar parameters ($T_{\rm eff}$, $\log_{10} g$, and $\log_{10} \mathscr{L}$) from light curve data using machine learning. Specifically, we compare two pathways as indicated by the top and bottom pathways in Figure \ref{fig1}. For both pathways, we first process the TESS light curve data into a format more suitable for machine learning analysis. We first normalize the light curves by their median fluxes and then transform each raw light curve into a Lomb–Scargle periodogram \citep{lomb1976, scargle1982}, a standard tool for analyzing periodic signals in unevenly sampled data \citep{vanderplas2018}. Each periodogram spans frequencies from 0.04 to 250 Hz with increments of 0.04 Hz (excluding the upper boundary), yielding 6,249 frequency samples of amplitude power per star. The amplitudes are then normalized by their power spectral density. After obtaining the periodogram representations for each light curve, we distinguish between the top and bottom pathways by whether we further modify the representations to input into neural networks.

In the bottom pathway of Figure \ref{fig1}, we standardize the periodogram before using it directly as input to the neural network. Specifically, we apply a base-10 logarithm to the amplitude values and then rescale them to the range [0,1]. The resulting log-scaled, normalized periodogram serves directly as input to the neural network. Formally, our prediction task can be expressed as:
\begin{equation}
    (T_{\rm eff}, \log_{10} g, \log_{10} \mathscr{L}) = f_{\theta}(\mathbf{x}),
\end{equation}
where $\mathbf{x}$ represents the log-scaled, normalized periodogram input, and $f_{\theta}$ is a neural network with learnable parameters $\theta$ that maps this periodogram information directly to the stellar parameters.

In the top pathway of Figure \ref{fig1}, we fit each periodogram to a red noise model, and the resulting four fitted and normalized parameters serve as input to the neural network. As discussed in Section \ref{sec:intro}, prior studies of OB stars have shown that the low-frequency red noise component of their periodograms carries stellar information \citep{blomme2011, bowman2019, bowman2020, anders2023}. Following these studies, we model the red noise of each periodogram with the function
\begin{equation}
    \alpha(\nu) = \frac{\alpha_0}{1+\left ( \frac{\nu}{\nu_{\rm char}}\right)^\gamma } + C_W,
    \label{eq3}
\end{equation}
where $\alpha_0$ is the amplitude at zero frequency, $\nu_{\rm char}$ is the characteristic frequency, $\gamma$ is the logarithmic amplitude gradient, and $C_W$ is a frequency-independent white noise term. Variations in the parameters $\alpha_0$, $\nu_{\rm char}$, and $\gamma$ have been shown to be potentially correlated with stellar parameters, implying that the observed red noise is of stellar rather than instrumental origin. We extract these four fitted parameters ($\alpha_0$, $\nu_{\rm char}$, $\gamma$, $C_W$), scale them to a uniform range [0,1], and use them as inputs for our machine learning task. 

The choice of model architecture matches the input representation. For the top pathway, having just four inputs representing the fitted red noise parameters makes a multilayer perceptron (MLP) a natural choice. An MLP is a feedforward neural network that maps inputs to outputs through fully connected layers \citep{rumelhart1986}. For the bottom pathway, the full periodogram is a high-dimensional and structured input, which is better suited to a convolutional neural network (CNN). CNNs apply sliding filters to detect localized or repeated patterns and are effective for structured one-dimensional sequences such as periodograms \citep{lecun1989}. The complementary strengths of these models motivate our comparative design: MLPs are efficient for learning direct relationships from low-dimensional representations, while CNNs are better suited for extracting complex patterns from higher-dimensional structured data. By comparing the two pathways, we assess whether the full periodogram contains predictive information beyond what is captured by the fitted parameters in Eq. \ref{eq3}.

Finally, in both pathways, we adopt a common training setup. We split our 285 O star light curves into two groups, with $80\%$ of the data for training and $20\%$ of the data for testing. The models are trained to minimize mean squared error (MSE) using the Adam optimizer with an initial learning rate of 1e-3. The learning rate is reduced by a factor of 0.5 if no improvement is observed for 15 (bottom pathway/CNN) and 20 (top pathway/MLP) consecutive epochs, and early stopping is applied if the test loss does not improve for 50 epochs. Rectified linear unit (ReLU) \citep{agarap2018} activations are used throughout for nonlinearity. These choices provide a consistent baseline across both the top and bottom pathways in Figure \ref{fig1}, and the best performing architecture details and performance are presented in Section~\ref{results}.

\begin{figure*}
    \centering
    \gridline{\fig{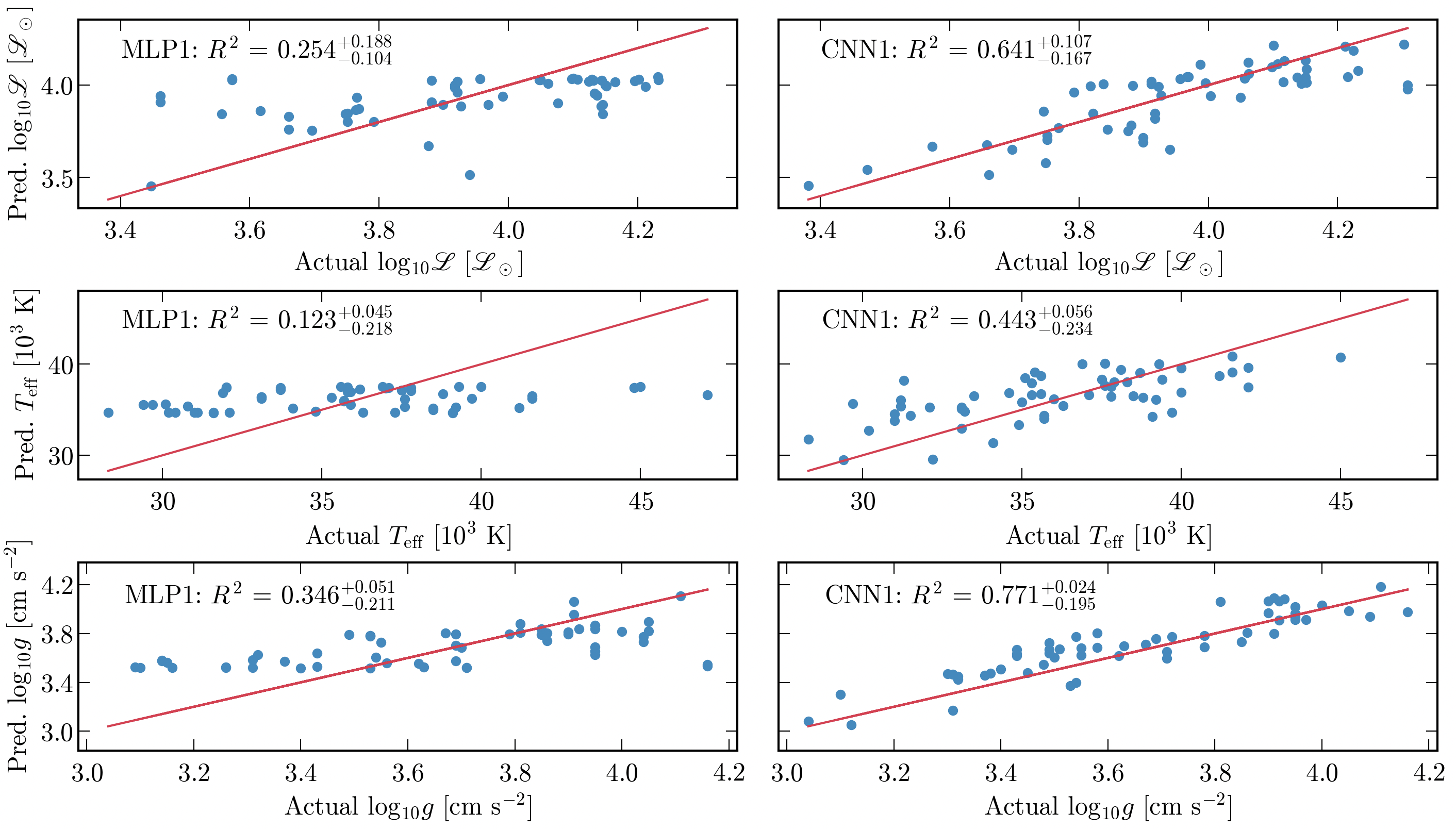}{0.95\textwidth}{(a)}} 
    \gridline{\fig{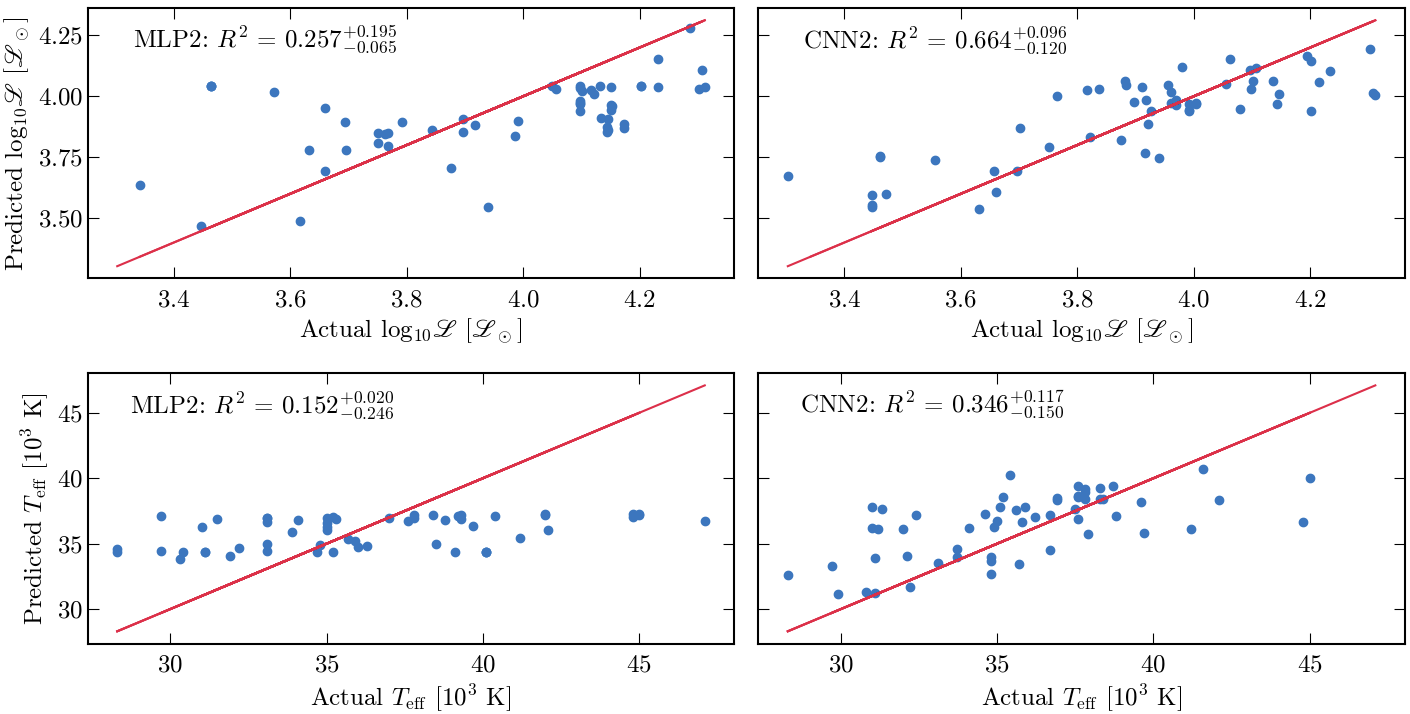}{0.95\textwidth}{(b)}} 
    \caption{Scatter plots in both panels compare actual and predicted values for spectroscopic luminosity (top rows), effective temperature (second rows), and surface gravity for MLP1/CNN1 (bottom row of Panel (a)). Panel (a) presents results from MLP1 and CNN1 models, while panel (b) presents results from MLP2 and CNN2 models. The red line in each plot indicates perfect agreement between predicted and actual values, serving as a reference for prediction accuracy.}
    \label{fig2}
\end{figure*}

\section{Results}
\label{results} 

\subsection{MLP on fitted red noise parameters}
For the top pathway of Figure \ref{fig1}, we systematically explored MLP architectures with 1-3 hidden layers and 64-512 neurons per layer (see details in Appendix \ref{appendix} and results in Tables \ref{appendixtable1} and \ref{appendixtable2}). The best performing MLP architectures, based on the average of the median $R^2$ performances in $T_{\rm eff}$ and $\log_{10} \mathscr{L}$ over 29 random seed runs (details in Appendix \ref{appendix}), are summarized in Table \ref{table1}. MLP1 predicts normalized ($T_{\rm eff}$, $\log_{10} g$) with three hidden layers of 256, 512, and 128 neurons, respectively, totaling 198,786 trainable parameters, and $\mathscr{L}$ is subsequently derived using Eq. \ref{eq1}. MLP2 predicts normalized ($T_{\rm eff}$, $\log_{10} \mathscr{L}$) with three hidden layers of 512, 256, and 128 neurons, respectively, totaling 167,042 trainable parameters.

We plot the performance of MLP1 and MLP2 in Figure \ref{fig2}. In the left panels of Figure~\ref{fig2}a, MLP1 shows limited effectiveness overall. For predicting $\log_{10} \mathscr{L}$, the upper left panel yields $R^2 = 0.254_{-0.104}^{+0.188}$, indicating that predictions capture only about $25.4\%$ of the variance in the observed values, with poor alignment to the ideal prediction line (shown in red). Predictions of $T_{\rm eff}$ perform worse, with $R^2 = 0.123_{-0.218}^{+0.045}$ (middle left panel). Predictions of $\log_{10} g$ achieve marginally better results, with $R^2 = 0.346_{-0.211}^{+0.051}$ (bottom left panel).

The left panels of Figure~\ref{fig2}b present results from MLP2. Performance is marginally improved relative to MLP1, with $R^2 = 0.257_{-0.065}^{+0.195}$ for $\log_{10} \mathscr{L}$ and $R^2 = 0.152_{-0.246}^{+0.020}$ for $T_{\rm eff}$. Because MLP2 performs slightly better overall, we adopt it for visualization in Figure~\ref{fig3}, and the details for choosing a representative random seed are in Appendix \ref{appendix}. When combining predictions of $T_{\rm eff}$ and $\log_{10} \mathscr{L}$ into the spectroscopic H–R diagram, the MLP2 predictions (purple squares) cluster around discrete groups, in contrast to the broader variance of the observed data (grey points). These results indicate that while the fitted red noise parameters identified in previous correlation studies (Section~\ref{sec:intro}) retain some predictive signal, their utility for accurately predicting stellar parameters is limited.

\subsection{CNN on full periodogram}

\begin{figure*}
    \centering
    \includegraphics[width=0.85\linewidth]{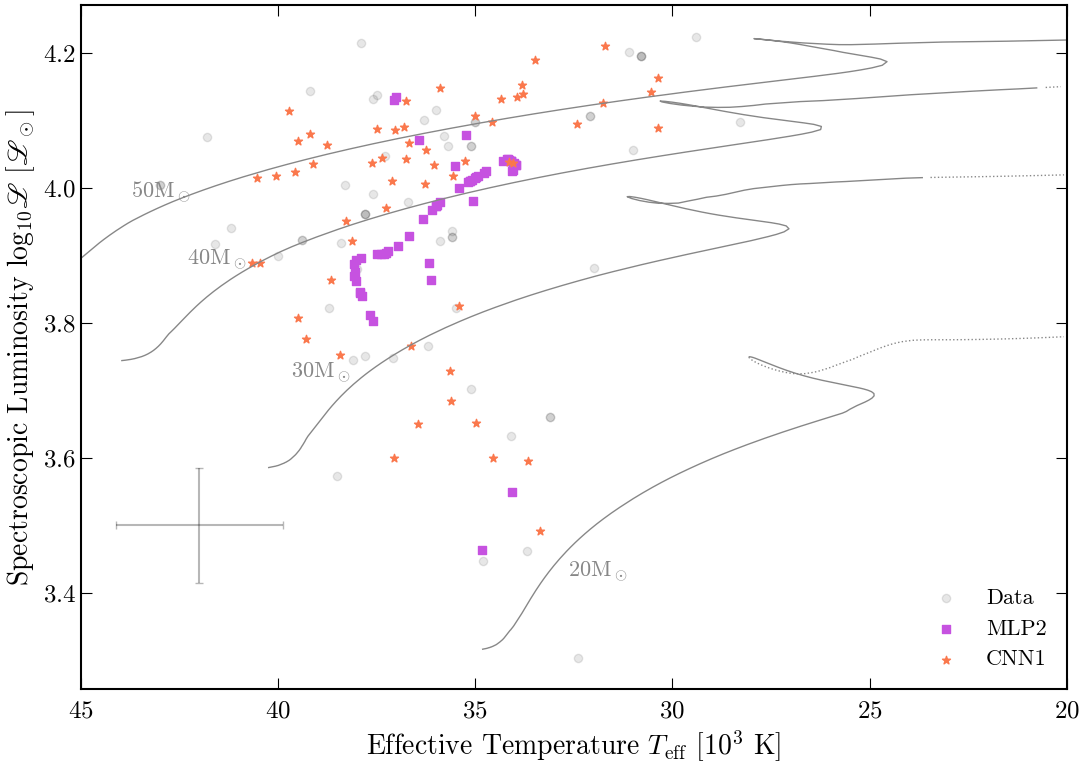}
    \caption{Spectroscopic H-R diagram comparing stellar parameter predictions from the most effective MLP and CNN models, specifically MLP2 (purple squares) and CNN1 (orange stars), against observed spectroscopic measurements of the stellar parameters (grey circles). The diagram plots effective temperature (in units of $10^3$ K) versus spectroscopic luminosity (in units of $\mathscr{L}_{\odot}$). Theoretical stellar evolutionary tracks from MIST (MESA Isochrones and Stellar Tracks; \citet{choi2016}) are overlaid, showing the predicted evolutionary pathways for stars with initial masses of 20--50 $M_{\odot}$. The solid portions of each track indicate the main-sequence phase, while dotted segments represent post-main-sequence phases. The representative error bar in the lower left indicates the mean prediction uncertainty of the CNN1 model relative to observational data.}
    \label{fig3}
\end{figure*}

For the bottom pathway of Figure \ref{fig1}, we explored a range of CNN configurations, varying the number of convolutional layers (1–3), fully connected layers (1–2), neurons per fully connected layer (64, 128), and channels (16, 32, 64) (see details in Appendix \ref{appendix} and results in Tables~\ref{appendixtable5} and~\ref{appendixtable6}). The best performing CNN architectures are summarized in Table~\ref{table1}. CNN1 predicts ($T_{\rm eff}$, $\log_{10} g$) with three convolutional layers followed by a fully connected layer, totaling 6,439,810 trainable parameters. Specifically, the first convolutional layer transforms the single-channel input periodogram into 64 feature maps using filters (kernel size=5, padding=2). The second and third convolutional layers maintain 64 feature maps with the same filter specifications. Each convolutional layer is followed by ReLU activation and max pooling (kernel size=2) to reduce dimensionality. The processed data then passes through a fully connected layer with 128 nodes. After predicting ($T_{\rm eff}$, $\log_{10} g$), $\mathscr{L}$ is calculated using Eq. \ref{eq1}, similar to MLP1. CNN2, on the other hand, directly predicts ($T_{\rm eff}$, $\log_{10} \mathscr{L}$), and the best performing CNN2 architecture is the same as CNN1, also totaling 6,439,810 trainable parameters. 

Both CNN1 and CNN2 trained on full periodogram inputs achieve superior performance in predicting $\log_{10} \mathscr{L}$ and $T_{\rm eff}$ compared to their respective MLP1 and MLP2 architectures that take the four fitted parameters $\alpha_0$, $\nu_{\rm char}$, $\gamma$, and $C_{\rm W}$ as input. As shown in the upper right panels of Figure~\ref{fig2}a and Figure~\ref{fig2}b, CNN1 and CNN2 achieve $R^2$ scores of $0.641_{-0.167}^{+0.107}$ and $0.664_{-0.120}^{+0.096}$, respectively, for predicting $\log_{10} \mathscr{L}$. These values represent a substantial improvement over the corresponding MLP models. Additionally, as shown in the middle right panel of Figure~\ref{fig2}a and lower right panel of Figure~\ref{fig2}b, CNN1 and CNN2 achieve $R^2$ scores of $0.443_{-0.234}^{+0.056}$ and $0.346_{-0.150}^{+0.117}$, respectively, for predicting $T_{\rm eff}$, which also outperforms their corresponding MLPs. CNN1 further achieves strong performance for predicting $\log_{10} g$ (lower right panel of Figure~\ref{fig2}a), with an $R^2$ value of $0.771_{-0.195}^{+0.024}$, which also surpasses the best MLP1 results. 

Notably, the superior performance of the CNN models is not simply due to its larger parameter count. To confirm this, we tested various MLP architectures ranging from 6-89 million trainable parameters using the standardized periodogram as input and found that despite having comparable to significantly more trainable parameters, the MLP architectures still underperformed CNNs (see Tables \ref{appendixtable3} and \ref{appendixtable4}). This indicates that not only is using the entire periodogram as input important, but also that the CNN architecture is specifically needed to extract the most information for stellar parameter prediction. Additionally, the performance of the CNN models is consistent across observations. To evaluate this, we examined stars observed in multiple TESS sectors and compared their predicted parameters across sectors. For CNN1, the mean per-star standard deviation was 0.57$\times10^3$\,K in $T_{\rm eff}$ and 0.0320\,dex in $\log_{10} g$, with mean ranges of 1.35$\times10^3$\,K and 0.0750\,dex, respectively. For CNN2, the corresponding values for $T_{\rm eff}$ and $\log_{10} \mathscr{L}$ were 0.33$\times10^3$\,K and 0.0156\,dex (mean standard deviations), and 0.79$\times10^3$\,K and 0.0373\,dex (mean ranges), respectively. These results indicate that predictions for the same star across different sectors are stable, with variations smaller than the typical prediction scatter visible in Figure \ref{fig2}, and therefore unlikely to reflect systematic differences between sectors.

Overall, CNN1 achieves the strongest performance for the joint predictions of $T_{\rm eff}$ and $\log_{10} \mathscr{L}$. In Figure \ref{fig3}, we present the spectroscopic H–R diagram by plotting the predictions of a representative random seed for $T_{\rm eff}$ versus $\log_{10} \mathscr{L}$ (see details for selecting representative random seed in Appendix \ref{appendix}). CNN1 predictions (orange stars) capture the variance of the data (grey circles) in both $T_{\rm eff}$ and $\log_{10} \mathscr{L}$ better than the MLP2 predictions (purple squares). This variance is astrophysically meaningful, as it reflects the spread of massive stars across different evolutionary phases, consistent with the MIST tracks (20–50 $M_\odot$) overlaid in the diagram (grey curves). To quantify the performance difference between MLP2 and CNN1, we calculate a multivariate $R^2$ that measures the fraction of total variance explained in the 2-D parameter space, accounting for the joint relationship between $T_{\rm eff}$ and $\log_{10} \mathscr{L}$ rather than treating each parameter independently. This metric uses the Euclidean distance between predicted and true values in the 2-D space, giving CNN1 a multivariate $R^2$ of 0.424 compared to MLP2's -0.060, indicating that CNN1 explains $42.4\%$ of the total variance of both parameters, while MLP2 performs worse than simply predicting the mean values. This confirms that this CNN architecture, when trained on the full periodogram, retains more stellar information than the MLP architecture trained on the fitted red noise parameters.

\section{Discussion}
\label{discussion}

This work demonstrates that representations of O star light curves contain sufficient information to predict physically meaningful stellar parameters. We compared two pathways: MLPs trained on fitted red noise parameters and CNNs trained on periodograms. Both approaches recover predictive signal, but our results show clearly that the CNNs trained on full periodograms retain substantially more information than MLPs trained on fitted red noise parameters alone. This outcome is astrophysically meaningful because the low-frequency red noise variability of massive stars is thought to arise from internal stellar processes such as internal gravity waves, subsurface convection, and wind instabilities, all of which are linked to stellar structure and evolution, but previous work has only identified correlations. By demonstrating predictive capability rather than just correlations, our study advances prior work and provides a framework for extracting this information from light curves alone. We focus on the spectroscopic H–R diagram rather than the classic H–R diagram because spectroscopic luminosity 1) does not require distance estimates and 2) depends on surface gravity, which changes as a star evolves, making the spectroscopic H–R diagram able to more clearly distinguish between different evolutionary states of massive stars.

Our strongest result highlighting these conclusions comes from predicting spectroscopic luminosity ($\log_{10} \mathscr{L}$). CNNs trained on periodograms achieve $R^2 = 0.641_{-0.167}^{+0.107}$ (CNN1) and $0.664_{-0.120}^{+0.096}$ (CNN2), far exceeding the performance of MLPs trained on fitted red noise parameters. Predictions of effective temperature ($T_{\rm eff}$) follow the same trend, with CNN1 reaching $R^2 = 0.443_{-0.234}^{+0.056}$ and CNN2 $R^2 = 0.346_{-0.150}^{+0.117}$. Although these values are more modest, they demonstrate that light curve data does contain some recoverable information about $T_{\rm eff}$ even for O stars, a population that is not only intrinsically more complex, but also represented in this paper by a significantly smaller dataset than prior studies of predicting $T_{\rm eff}$ from light curves of cooler stars \citep{miller2015, blancato2022}. As a byproduct of predicting $\log_{10} \mathscr{L}$ well, CNN1 also predicts surface gravity with $R^2 = 0.771_{-0.195}^{+0.024}$, reflecting the close relationship among these stellar parameters and underscoring that these O star light curves encode detailed evolutionary information.

The performance differences between the two pathways in predicting $T_{\rm eff}$ and $\log_{10} \mathscr{L}$ are clearest in the spectroscopic H–R diagram (Figure~\ref{fig3}). The MLP pathway is overconfident in its predictions, producing clusters on the diagram that fail to reproduce the observed variance in $T_{\rm eff}$ and $\log_{10} \mathscr{L}$. The CNN pathway, in contrast, recovers the broader spread of massive stars in this parameter space, aligning more closely with both the observed distribution and the theoretical evolutionary tracks. The multivariate $R^2$ across ($T_{\rm eff}$, $\log_{10} \mathscr{L}$) confirms this difference: CNN1 achieves 0.424, while MLP2 scores –0.060, worse than predicting the mean. Capturing this variance is astrophysically significant, since it reflects the diversity of evolutionary states among massive stars.

Some methodological considerations merit further discussion. As shown in all the subfigures of Figure \ref{fig2}, our models tend to predict values clustering around the mean, possibly overlooking the nuances at both the low and high ends of the spectrum. This pattern is consistent with the non-uniform composition of our sample reported in Section \ref{sec:data} (19 early, 53 mid, 34 late; 53 dwarfs, 16 giants, 37 supergiants), which yields fewer training examples at more extremes. This behavior is analogous to the class imbalance issue in classification tasks, where models are biased towards more frequently represented categories. Future work could explore strategies to enhance the variance in predictions via methods such as loss re-weighting, stratified train/test splits, and inverse-frequency sampling, potentially improving the model's ability to capture the full range of stellar parameter values. 

Additionally, the choice between using periodograms or light curves as input presents another important consideration. While periodograms transformed from light curves effectively capture periodic phenomena crucial for stellar characterization, they still represent a dimension reduction of the original light curve data. This transformation might lose potentially valuable information present in the raw time series. However, working directly with light curves presents its own challenges, including increased noise and the need for more complex input processing for machine learning models. The trade-offs between these approaches warrant further investigation as we seek to maximize the stellar information we can extract from time-series photometric data.

A further methodological consideration concerns the scope of our sample. In this paper, we deliberately restrict our analysis to O stars. Although the IACOB database also includes early B stars analyzed with the same {\sc iacob-broad} and {\sc iacob-gbat} tools, the O star subset has been derived through a fully homogeneous pipeline and calibration scheme that has been extensively vetted for internal consistency. Incorporating B stars at this stage would introduce slight systematic inhomogeneities, as the adopted model grids and parameter calibrations differ in detail from those used for O stars, potentially biasing the machine learning inference. In addition, the inclusion of B stars would numerically dominate the training set and shift the model’s focus toward their parameter regime, effectively treating O stars as outliers. Finally, photometric phenomenology (e.g., SPB/$\beta$ Cep pulsations common in B stars) differs from O star variability, adding an additional layer of complexity that we prefer to decouple in this first analysis. Extending the framework presented here to early B-type stars using a consistently calibrated dataset will be a natural next step toward a unified treatment of massive stars.

In conclusion, our work establishes that (1) O star light curves contain sufficient information to predict stellar parameters, and (2) CNNs trained on full periodograms recover substantially more stellar information than MLPs trained on fitted red noise parameters. As surveys like LSST collect unprecedented volumes of light curve data without accompanying spectroscopy, the ability to infer stellar parameters from light curves becomes increasingly valuable. Our approach enables the preliminary characterization of massive stars at scale and helping to prioritize targets for spectroscopic follow-up. Extending this methodology to systematically include a broader range of stellar types will further enhance its utility in advancing our understanding of stellar physics in the era of large-scale surveys.

\section{Acknowledgements}
\begin{acknowledgments}
R.Z. acknowledges the support and hospitality from the Simons Foundation through the predoctoral program at the Center for Computational Astrophysics, Flatiron Institute. The computational resources and services used in this work were provided by facilities supported by the Scientific Computing Core at the Flatiron Institute, a division of the Simons Foundation. The Center for Computational Astrophysics at the Flatiron Institute is supported by the Simons Foundation. G.H. acknowledges support from the State Research Agency (AEI) of the Spanish Ministry of Science and Innovation (MICIN) under grants PID2021-122397NB-C21 / PID2022-136640NB-C22, and the European Regional Development Fund, FEDER 10.13039-501100011033. Based on observations made with the Nordic Optical Telescope, operated by NOTSA, and the Mercator Telescope, operated by the Flemish Community, both at the Observatorio del Roque de los Muchachos (La Palma, Spain) of the Instituto de Astrofísica de Canarias. 
\end{acknowledgments}

\appendix
\renewcommand{\thetable}{A.\arabic{table}}
\setcounter{table}{0}
\section{Implementation details and architecture search}
\label{appendix}

We performed an extensive architecture search for our task of predicting stellar parameters using two types of input representations: (1) the four fitted red noise parameters ($\alpha_0$, $\gamma$, $\nu_{\rm char}$, and $C_{\rm W}$), and (2) the full periodograms. Each model was trained for up to 500 epochs with early stopping using a patience of 50 epochs. Epoch counts and training times are rounded to the nearest hundredth, and $R^2$ values are reported to four decimal places. We ran 29 instances of each architecture, each with a fixed but different random seed. All relevant random seeds were set to ensure full reproducibility of the regression pipeline, including data splitting and model training. Specifically, we set 29 different seeds for Python’s \texttt{random} module, \texttt{NumPy}, and \texttt{PyTorch} (both CPU and GPU), and configured \texttt{PyTorch}'s backend for deterministic behavior by setting \texttt{torch.backends.cudnn.deterministic = True} and disabling benchmarking. For data splitting, we used a \texttt{torch.Generator} with the same 29 fixed seeds to ensure consistent train/test splits across runs. 

For each architecture, we summarize the 29 runs by reporting the median-performing seed, ranked by the average of $T_{\rm eff}$ and $\log_{10} \mathscr{L}$ $R^2$ values. Figure~\ref{fig2} plots the median seeds of the best performing architectures ranked by this metric. To quantify the uncertainty in all $R^2$ estimates, we compute the 10th and 90th percentile values across these 29 runs as the lower and upper bounds, respectively. These percentiles (q10 and q90) capture the range within which the central $80\%$ of model performances fall, providing a measure of variability due to randomness in initialization and data splitting. For the spectroscopic H–R diagram in Figure \ref{fig3}, we select a representative random seed using a three-step procedure: (1) for each seed and architecture, we calculate the average of the $T_{\rm eff}$ and $\log_{10} \mathscr{L}$ $R^2$ values; (2) for each seed, we then compute the mean of these averages across all four architectures; and (3) we select the seed with the median performance among the 29. Using this single seed ensures that the same test dataset is applied consistently to both the true data values and the model predictions. 

The remainder of this appendix summarizes the explored architectures, their configurations, and the results obtained across multiple random seed runs. For models trained on the four fitted red noise parameters, we systematically evaluated multiple MLP configurations, testing 29 random seeds for each architecture; Tables \ref{appendixtable1} and \ref{appendixtable2} summarize these results, with the best-performing MLP1 and MLP2 architectures highlighted in bold. We also tested MLPs on the full periodogram input (Tables \ref{appendixtable3} and \ref{appendixtable4}), running 10 seeds per architecture; however, these models showed poor convergence and generalization, so no single architecture was selected. These findings motivated the use of CNNs, for which we again systematically explored multiple configurations with 29 seeds each; the performance summaries are presented in Tables \ref{appendixtable5} and \ref{appendixtable6}, with the best-performing CNN1 and CNN2 architectures highlighted in bold.

\begin{deluxetable}{ccccccc}
\tablecaption{Performance summary of candidate MLP1 architectures
\label{appendixtable1}}
\tablewidth{0pt}
\tablehead{
\colhead{Architecture} & \colhead{\shortstack{Average \\ Epochs}} & \colhead{\shortstack{Average Train \\ Time (min)}} & \colhead{\shortstack{Trainable \\ Parameters}} & \colhead{\shortstack{$T_{\rm eff}$ $R^2$ \\ Median}} & \colhead{\shortstack{$\log_{10} g$ $R^2$ \\ Median}} & \colhead{\shortstack{$\log_{10} \mathscr{L}$ $R^2$ \\ Median}}
}
\startdata
{[}128{]} & 133.66 & 0.28 & 898 & 0.0906 & 0.2693 & 0.2818\\
{[}256{]} & 109.38 & 0.22 & 1794 & 0.0960 & 0.2665 & 0.2799\\
{[}512{]} & 105.03 & 0.21 & 3586 & 0.0909 & 0.2705 & 0.2846\\
{[}256, 128{]} & 103.69 & 0.23 & 34434 & 0.0844 & 0.2762 & 0.2965\\
{[}128, 256{]} & 90.24 & 0.21 & 34178 & 0.0789 & 0.2705 & 0.2889\\
{[}256, 256{]} & 105.72 & 0.25 & 67586 & 0.0863 & 0.2791 & 0.3141\\
{[}256, 128, 64{]} & 95.17 & 0.24 & 42562 & 0.0756 & 0.2755 & 0.3123\\
{[}512, 256, 128{]} & 114.79 & 0.31 & 167042 & 0.0948 & 0.2915 & 0.3059\\
{[}128, 256, 128{]} & 113.93 & 0.27 & 66818 & 0.0746 & 0.2724 & 0.3150\\
\textbf{[256, 512, 128]} & \textbf{123.41} & \textbf{0.33} & 
\textbf{198786} & \textbf{0.0863} & \textbf{0.2912} & \textbf{0.3208}\\
{[}256, 256, 256{]} & 120.45 & 0.33 & 133378 & 0.0828 & 0.2886 & 0.3219\\
\enddata
\end{deluxetable}

\begin{deluxetable}{ccccccc}
\tablecaption{Performance summary of candidate MLP2 architectures 
\label{appendixtable2}}
\tablewidth{0pt}
\tablehead{
\colhead{Architecture} & \colhead{\shortstack{Average \\ Epochs}} & \colhead{\shortstack{Average Train \\ Time (min)}} & \colhead{\shortstack{Trainable \\ Parameters}} & \colhead{\shortstack{$T_{\rm eff}$ $R^2$ \\ Median}} & \colhead{\shortstack{$\log_{10} \mathscr{L}$ $R^2$ \\ Median}}
}
\startdata
{[}128{]} & 189.0 & 0.38 & 898 & 0.0882 & 0.2931\\
{[}256{]} & 168.52 & 0.36 & 1794 & 0.0900 & 0.2956\\
{[}512{]} & 134.07 & 0.28 & 3586 & 0.0983 & 0.3027\\
{[}256, 128{]} & 127.48 & 0.30 & 34434 & 0.0706 & 0.3141\\
{[}128, 256{]} & 126.38 & 0.31 & 34178 & 0.0786 & 0.3189\\
{[}256, 256{]} & 122.55 & 0.31 & 67586 & 0.0735 & 0.3052\\
{[}256, 128, 64{]} & 148.52 & 0.40 & 42562 & 0.0788 & 0.3165\\
\textbf{[512, 256, 128]} & \textbf{137.07} & \textbf{0.38} & 
\textbf{167042} & \textbf{0.0866} & \textbf{0.3464}\\
{[}128, 256, 128{]} & 134.17 & 0.36 & 66818 & 0.0854 & 0.3195\\
{[}256, 512, 128{]} & 122.52 & 0.35 & 198786 & 0.0893 & 0.3193\\
{[}256, 256, 256{]} & 130.28 & 0.40 & 133378 & 0.0781 & 0.3317\\
\enddata
\end{deluxetable}

\begin{deluxetable}{ccccccc}
\tablecaption{Performance summary of candidate MLP architectures predicting $T_{\rm eff}$ and $\log_{10} g$ from full periodogram inputs \label{appendixtable3}}
\tablewidth{0pt}
\tablehead{
\colhead{Architecture} & \colhead{\shortstack{Average \\ Epochs}} & \colhead{\shortstack{Average Train \\ Time (min)}} & \colhead{\shortstack{Trainable \\ Parameters}} & \colhead{\shortstack{$T_{\rm eff}$ $R^2$ \\ Median}} & \colhead{\shortstack{$\log_{10} g$ $R^2$ \\ Median}} & \colhead{\shortstack{$\log_{10} \mathscr{L}$ $R^2$ \\ Median}}
}
\startdata
{[}1024{]} & 336.9 & 0.08 & 6402050 & 0.2760 & 0.4676 & 0.3358\\
{[}2048{]} & 393.1 & 0.13 & 12804098 & 0.2772 & 0.5243 & 0.3372\\
{[}4096{]} & 330.2 & 0.16 & 25608194 & 0.2631 & 0.4844 & 0.3188\\
{[}8192{]} & 393.8 & 0.32 & 51216386 & 0.2854 & 0.5100 & 0.3254\\
{[}8192, 2048{]} & 252.5 & 0.27 & 67983362 & 0.1063 & 0.3745 & 0.2780\\
{[}8192, 1024{]} & 333.9 & 0.31 & 59591682 & 0.2557 & 0.5218 & 0.3410\\
{[}4096, 1024{]} & 200.2 & 0.11 & 29797378 & -0.0007 & 0.1447 & 0.1544\\
{[}2048, 512{]} & 169.5 & 0.06 & 13850114 & -0.0003 & -0.0001 & -0.0008\\
{[}8192, 2048, 512{]} & 108.5 & 0.12 & 69029378 & -0.0002 & -0.0005 & -0.0010\\
{[}4096, 2048, 512{]} & 209.7 & 0.13 & 35040770 & -0.0001 & 0.36023 & 0.3558\\
{[}8192, 4096, 1024{]} & 61.3 & 0.08 & 88955906 & -0.0029 & -0.0092 & -0.0074\\
\enddata
\end{deluxetable}

\begin{deluxetable}{ccccccc}
\tablecaption{Performance summary of candidate MLP architectures predicting $T_{\rm eff}$ and $\log_{10} \mathscr{L}$ from full periodogram inputs
\label{appendixtable4}}
\tablewidth{0pt}
\tablehead{
\colhead{Architecture} & \colhead{\shortstack{Average \\ Epochs}} & \colhead{\shortstack{Average Train \\ Time (min)}} & \colhead{\shortstack{Trainable \\ Parameters}} & \colhead{\shortstack{$T_{\rm eff}$ $R^2$ \\ Median}} & \colhead{\shortstack{$\log_{10} \mathscr{L}$ $R^2$ \\ Median}}
}
\startdata
{[}1024{]} & 310.5 & 0.08 & 6402050 & 0.2163 & 0.3377\\
{[}2048{]} & 388.2 & 0.13 & 12804098 & 0.2291 & 0.3569\\
{[}4096{]} & 305.6 & 0.14 & 25608194 & 0.2016 & 0.3358\\
{[}8192{]} & 343.0 & 0.28 & 51216386 & 0.1875 & 0.3349\\
{[}8192, 2048{]} & 279.9 & 0.30 & 67983362 & 0.1209 & 0.3307\\
{[}8192, 1024{]} & 334.8 & 0.31 & 59591682 & 0.1766 & 0.3721\\
{[}4096, 1024{]} & 251.7 & 0.14 & 29797378 & 0.0662 & 0.3179\\
{[}2048, 512{]} & 266.4 & 0.09 & 13850114 & 0.1420 & 0.3897\\
{[}8192, 2048, 512{]} & 170.2 & 0.18 & 69029378 & -0.0002 & -0.0026\\
{[}4096, 2048, 512{]} & 243.2 & 0.15 & 35040770 & 0.0917 & 0.4370\\
{[}8192, 4096, 1024{]} & 86.2 & 0.12 & 88955906 & -0.0029 & -0.0061\\
\enddata
\end{deluxetable}

\begin{deluxetable}{ccccccc}
\tablecaption{Performance summary of candidate CNN1 architectures
\label{appendixtable5}}
\tablewidth{0pt}
\tablehead{
\colhead{Architecture} & \colhead{\shortstack{Average \\ Epochs}} & \colhead{\shortstack{Average Train \\ Time (min)}} & \colhead{\shortstack{Trainable \\ Parameters}} & \colhead{\shortstack{$T_{\rm eff}$ $R^2$ \\ Median}} & \colhead{\shortstack{$\log_{10} g$ $R^2$ \\ Median}} & \colhead{\shortstack{$\log_{10} \mathscr{L}$ $R^2$ \\ Median}}
}
\startdata
1conv-16ch-64fc & 79.14 & 0.43 & 3199266 & -0.0080 & -0.0030 & -0.0033\\
1conv-32ch-64fc & 128.59 & 0.88 & 6398338 & -0.0024 & -0.0015 & -0.0032\\
1conv-32ch-128fc & 156.10 & 1.17 & 12796482 & 0.0478 & 0.1756 & 0.1133\\
1conv-64ch-128fc & 156.69 & 1.80 & 25592578 & -0.0005 & -0.0001 & -0.0005\\
2conv-16ch-64fc & 115.90 & 0.87 & 1601074 & 0.1898 & 0.4831 & 0.5405\\
2conv-32ch-64fc & 106.52 & 0.82 & 3204514 & 0.2306 & 0.5491 & 0.5692\\
2conv-32ch-128fc & 108.45 & 0.92 & 6403682 & 0.2552 & 0.6188 & 0.6000\\
2conv-64ch-128fc & 113.72 & 1.45 & 12817218 & 0.2873 & 0.6088 & 0.5954\\
2conv-32ch-256fc & 103.34 & 1.03 & 12802018 & 0.2840 & 0.5898 & 0.6130\\
3conv-16ch-64fc & 145.93 & 1.38 & 802626 & 0.3387 & 0.6559 & 0.6079\\
3conv-32ch-64fc & 136.90 & 1.18 & 1610178 & 0.3786 & 0.7067 & 0.6386\\
3conv-32ch-128fc & 128.52 & 1.19 & 3209858 & 0.3597 & 0.6899 & 0.6724\\
\textbf{3conv-64ch-128fc} & \textbf{137.10} & \textbf{1.92} & 
\textbf{6439810} & \textbf{0.4264} & \textbf{0.7369} & \textbf{0.6560}\\
2conv-32ch-128fc-64fc & 104.55 & 0.88 & 6411810 & 0.3067 & 0.6608 & 0.6349\\
2conv-32ch-256fc-128fc & 107.52 & 1.09 & 12834658 & 0.3166 & 0.6748 & 0.6276\\
2conv-64ch-128fc-64fc & 119.28 & 1.61 & 12825346 & 0.3297 & 0.7088 & 0.6628\\
\enddata
\end{deluxetable}

\begin{deluxetable}{cccccc}
\tablecaption{Performance summary of candidate CNN2 architectures
\label{appendixtable6}}
\tablewidth{0pt}
\tablehead{
\colhead{Architecture} & \colhead{\shortstack{Average \\ Epochs}} & \colhead{\shortstack{Average Train \\ Time (min)}} & \colhead{\shortstack{Trainable \\ Parameters}} & \colhead{\shortstack{$T_{\rm eff}$ $R^2$ \\ Median}} & \colhead{\shortstack{$\log_{10} \mathscr{L}$ $R^2$ \\ Median}}
}
\startdata
1conv-16ch-64fc & 86.52 & 0.51 & 3199266 & -0.0006 & -0.0007\\
1conv-32ch-64fc & 136.0 & 0.93 & 6398338 & -0.0023 & -0.0012\\
1conv-32ch-128fc & 141.0 & 1.19 & 12796482 & -0.0001 & -0.0001\\
1conv-64ch-128fc & 161.69 & 1.69 & 25592578 & -0.0016 & -0.0002\\
2conv-16ch-64fc & 110.52 & 0.85 & 1601074 & 0.1010 & 0.5491\\
2conv-32ch-64fc & 119.10 & 0.92 & 3204514 & 0.2409 & 0.6010\\
2conv-32ch-128fc & 122.59 & 1.03 & 6403682 & 0.2718 & 0.5933\\
2conv-64ch-128fc & 122.10 & 1.50 & 12817218 & 0.2780 & 0.6235\\
2conv-32ch-256fc & 103.83 & 1.04 & 12802018 & 0.2629 & 0.6128\\
3conv-16ch-64fc & 130.41 & 1.24 & 802626 & 0.2780 & 0.5967\\
3conv-32ch-64fc & 135.59 & 1.19 & 1610178 & 0.3327 & 0.6285\\
3conv-32ch-128fc & 127.72 & 1.16 & 3209858 & 0.3288 & 0.6540\\
\textbf{3conv-64ch-128fc} & \textbf{133.28} & \textbf{1.85} & 
\textbf{6439810} & \textbf{0.3621} & \textbf{0.6637}\\
2conv-32ch-128fc-64fc & 112.17 & 0.97 & 6411810 & 0.2700 & 0.6398\\
2conv-32ch-256fc-128fc & 107.59 & 1.08 & 12834658 & 0.2927 & 0.6462\\
2conv-64ch-128fc-64fc & 127.34 & 1.61 & 12825346 & 0.2770 & 0.6511\\
\enddata
\end{deluxetable}


\end{document}